\documentclass{article}
\usepackage{spconf,amsmath,graphicx}
\usepackage{multirow}
\usepackage[lined,ruled]{algorithm2e}
\usepackage{multibib}
\usepackage{subfig}
\usepackage{booktabs}
\usepackage{amssymb}
\usepackage{tikz}
\usepackage[utf8]{inputenc}
\usepackage[russian,english]{babel}

\usepackage[colorinlistoftodos]{todonotes}

\title{UserReg: A Simple but Strong Model for Rating Prediction}
\name{Haiyang Zhang$^{1}$ \qquad Ivan Ganchev$^{2,3,4*}$ \qquad Nikola S. Nikolov$^{4**}$ \qquad Mark Stevenson$^{1}$
\thanks{
This publication has been supported by the Bulgarian National Science Fund (BNSF) under the Grant No. KP-06-IP-CHINA/1.}
}
  
\address{\ninept$^{1}$ Department of Computer Science, The University of Sheffield, UK \\
      \ninept$^{2}$Department of Computer Systems, University of Plovdiv “Paisii Hilendarski”, Bulgaria\\
      \ninept$^{3}$Institute of Mathematics and Informatics - Bulgarian Academy of Sciences, Bulgaria\\
      \ninept$^{4}$ TRC*/CSIS**,University of Limerick, Ireland}

\begin{document}
	\ninept
	\maketitle
	\begin{abstract}
		Collaborative filtering (CF) has achieved great success in the field of recommender systems. 
		In recent years, many novel CF models, particularly those based on deep learning or graph techniques, have been proposed for a variety of recommendation tasks, such as rating prediction and item ranking. These newly published models usually demonstrate their performance in comparison to baselines or existing models in terms of accuracy improvements. However, others have pointed out that many newly proposed models are not as strong as expected and are outperformed by very simple baselines. 
		
	This paper proposes a simple linear model based on Matrix Factorization (MF), called {\it UserReg}, which regularizes users' latent representations with explicit feedback information for rating prediction. We compare the effectiveness of UserReg with three linear CF models that are widely-used as baselines, and with a set of recently proposed complex models that are based on deep learning or graph techniques. Experimental results show that UserReg achieves overall better performance than the fine-tuned baselines considered and is highly competitive when compared with other recently proposed models. We conclude that UserReg can be used as a strong baseline for future CF research.  
%
	\end{abstract}

		\vspace{-6pt}	
	\section{Introduction}
	\label{sec:intro}
		\vspace{-6pt}	

	Personalized recommendation techniques have become popular and are widely used in many applications, including e-commerce and online learning. They can be deployed to help users discover useful information and help merchants reach valuable target users. The most successful and widely deployed recommendation technique is collaborative filtering (CF) due to its effectiveness, efficiency and the fact that it can make recommendations based only on users' historical interactions and without requiring specific domain information \cite{funk2006netflix,koren2009matrix}. 
	Within CF, embedding based models have been the most popular ones for over a decade \cite{koren2009matrix,WorryingAnalysis2019,rendle2019difficulty}. 
	Matrix factorization (MF) is among the simplest and most widely used embedding based approaches \cite{koren2009matrix}. Numerous extensions to MF have been proposed to further improve prediction accuracy including (1) traditional enhancements, such as PMF \cite{2007probabilisticMF}, BiasedMF \cite{koren2009matrix} and SVD++ \cite{SVD++2008}, proposed a decade ago and widely considered as baselines in current research, and (2) hybrid models that make use of content information related to users/items for achieving further improvements \cite{ma2009RSTE,SoReg2011,guo2016TrustSVD,HERec2019,HeteMF2013,DSR2016,SemRec2015}.   	
%
	
	In recent years, embedding based models using deep learning techniques have become particular popular, following their success in multiple application domains such as computer vision and natural language processing. Most of these models reported significant improvements over baselines \cite{AutoRec2015,AutoSVD++2017,CF-NADE2016,Zhang2020IGMC,Graphrec2019}. However, some recent works have pointed out that models based on deep learning are not as strong as expected while also being computational complex \cite{WorryingAnalysis2019,rendle2019difficulty}. For instance, Rendle \textit{et al.} \cite{rendle2019difficulty} showed that numerous recommendation models published in top-level conferences between 2015 and 2019 are outperformed by long-established baselines, which were even 
	not the best ones. We also found that baseline results reported
in some recent work are suboptimal, which aligns with these findings. Dacrema \textit{et al.} \cite{WorryingAnalysis2019} analyzed a number of the recent neural network (NN) based models for top-n recommendation and found most of them failed to outperform simple baselines. 
	
	Recommendation problems can be cast as either rating prediction or item ranking (also known as top-n recommendation) task and most algorithms are designed for only one of these \cite{guo2016TrustSVD}. In this paper, we focus on the rating prediction task which predicts the value of unobserved ratings to recommend items with the highest estimated scores. 
	Our contributions can be summarized as follows:
	\begin{itemize}
		\item We propose a conceptually and computationally simple MF-based model, UserReg, that effectively makes use of user explicit feedback to regularize user representation in MF in order to improve recommendation accuracy. The experiments conducted on four widely used datasets demonstrate that our model outperforms popular fine-tuned baselines and can achieve state-of-the-art performance. In particular, it outperforms the majority of MF-based models that use additional content information; 
		\item We found UserReg more efficient in both training and recommendation compared to SVD++, which is considered as a strong baseline \cite{rendle2019difficulty,guo2016TrustSVD}. We conclude that UserReg can be used as a strong baseline for future development of CF approaches;  
		\item We found that the fine-tuned baseline results are better than those reported in some recent works under the same experimental settings and that several MF-based approaches were not compared against strong baselines (e.g. SVD++), and would be outperformed by them. 
	\end{itemize}
	
	\vspace{-5pt}	
	\section{Matrix Factorization for Rating Prediction}
	\vspace{-6pt}	
	A typical recommendation scenario involves a set of $m$ users $ U=\{u_{1},u_{2}...u_{m}\} $, a set of $n$ items $ I=\{i_{1},i_{2}...i_{n}\}$, and their interactions represented by a rating matrix $R \in \mathbb{R}_{m,n}$. The goal of MF is to map both users and items into the same low-rank latent factor space by approximating the observed ratings. The users and items are represented by a set of feature vectors, $P \in \mathbb{R}_{m, d}$ and $Q \in \mathbb{R}_{n, d}$, respectively, where $ d $ is the number of latent factors. The predicted rating  of item \textit{i} by user \textit{u} is calculated by their dot product, as: $\tilde{r}_{ui}=P_uQ_i^T$. The matrices $P$ and $Q$ are learnt by minimizing the regularized squared error on the set of observed ratings $(u, i) \in K$: 
	\begin{equation}\footnotesize
		\label{eq:objoriginal}
		O_{MF}=\sum_{(u, i)\in K}\Big((r_{ui} - \tilde{r}_{ui})^{2}+\lambda(\|P_{u}\|^{2}+\|Q_{i}\|^{2})\Big)
	\end{equation}
	where 
	{\small$\lambda(\|P_{u}\|^{2}+\|Q_{i}\|^{2})$} is the regularization term.
	
	Several enhancements to MF have been proposed and  are widely used as baselines for comparison with newly proposed models: 
	
	(1) Probabilistic matrix factorization (\textbf{PMF}) \cite{2007probabilisticMF} formulates MF from a probabilistic perspective, assuming that entries of $R \in \mathbb{R}_{m,n}$ are independent and normally distributed. The conditional distribution over the observed ratings is defined as:
	\begin{equation}\footnotesize
		\label{eq:PMF}
		p(r|P,Q, \sigma^2) = \prod_{u=1}^m\prod_{i=1}^{n}[\mathcal{N}(r_{u,i}|P_uQ_i^T, \sigma^2)]^{I_{u,i}}
	\end{equation}
	where $ \mathcal{N}(x|\mu , \sigma^2) $ is the probability density function of the Gaussian  distribution and $ I_{u,i} $ is a binary indicator to check if $ r_{u,i} $ is an observed rating.
	
	(2) \textbf{BiasedMF}  (also known as BiasedSVD) improves conventional MF by introducing bias information for users and items \cite{koren2009matrix}. 
	BiasedMF predicts rating values as:
	\begin{equation}\footnotesize
		\label{eq:BiasedMF}
		\tilde{r}_{ui}=P_uQ_i^T + \mu + b_{u}+b_{i}
	\end{equation}
	where $ \mu $ is the global average rating, and $ b_{u} $ and $ b_{i} $ indicate the user bias and item bias, respectively.
	
	(3) \textbf{SVD++} \cite{SVD++2008} extends BiasedMF, assuming that the rating of user \textit{u} on item \textit{i} is not only related to user $u$ and item $i$, but also to the items that $u$ has already rated. It incorporates implicit feedback information into BiasedMF:
	\begin{equation}\footnotesize
		\label{eq:SVD++}
		\tilde{r}_{ui}=(P_{u}+|R_{u}|^{- \frac{1}{2}}\sum_{j\in R_{u}}y_{j})Q_{i}^{T}+b_{ui}
	\end{equation}
	where $b_{ui}=\mu +b_{u}+b_{i}$ represents the bias term, $y_{j} \in \mathbb{R}_{d}$ represents the implicit influence of implicit feedback by user $u$, and $ R_{u} $ is the set of items user $u$ has rated.
	\vspace{-5pt}	
	\section{UserReg}
	The basic form of MF assumes that the rating $ r_{ui} $ is only based on the latent representation of user $u$ and item $i$. Some advanced forms of MF go one step further by assuming $ r_{ui} $ also relies on other context, e.g. users' implicit feedback in SVD++ \cite{SVD++2008}, users' social relations \cite{SoReg2011,guo2016TrustSVD,TrustMF2017}, or side information of users and items \cite{AutoSVD++2017,HERec2019}. Similar to the previous works, the assumptions behind the proposed model are that (1) $ r_{ui} $ is not only related to user $u$ and item $i$ but also to other items that user $u$ has shown interest in (e.g. items that received high ratings by user $u$ ); and (2) the latent feature vector of user $u$ should be close to the set of items of interest to that user. In line with these assumptions, the proposed model \textbf{UserReg} incorporates user explicit feedback into MF as a regularization term to constrain the MF objective function:
	\begin{equation}\footnotesize
		\label{eq:User-Reg}
		\begin{aligned}
			O_{UR}&=\frac{1}{2}\sum_{u \in U}\sum_{i\in R_{u}}(r_{ui}-\tilde{r}_{ui})^{2} \\
			&+\frac{\beta}{2} \sum_{u \in U}|| P_{u}-|I(u)|^{-1}\sum_{j\in I(u)}Q_{j}||^{2}_{F}\\
			&+\frac{\lambda}{2}(\sum_{u \in U}\|P_{u}\|^{2}_{F}+ \sum_{i \in I}\|Q_{i}\|^{2}_{F} + \sum_{u \in U}\|b_{u}\|^{2}+\sum_{i \in I}\|b_{i}\|^{2})
		\end{aligned}
	\end{equation}
	where $R_{u}$ is the set of items rated by user $u$, $I(u)$ is the set of items user $u$ is interested in (we use items with a rating higher than the average), $\beta > 0$  is a regularization parameter that controls the influence of user feedback, and $ \|\cdot\|^{2}_F $ is the Frobenius norm.
	The regularization term aims to minimize the distance ${\textstyle  \| P_{u}-|I(u)|^{-1}\sum_{j\in I(u)}Q_{j}||^{2}_{F} }$, such that the latent representation for user $ u $ is close to the average representation of items that $ u $ has shown interest in, {\small  $ P_{u} \approx |I(u)|^{-1}\sum_{j\in I(u)}Q_{j} $}. Therefore, the dot product of $ P_{u} $ and $ Q_{i} $ can be approximated as:
	\begin{equation}\footnotesize
		\label{eq:inner}
		\langle P_{u}, Q_{i}\rangle \approx \langle |I(u)|^{-1}\sum_{j\in I(u)}Q_{j},Q_{i} \rangle = 
		|I(u)|^{-1}\sum_{j\in I(u)} \langle Q_{j},Q_{i} \rangle
	\end{equation}
	That is, the predicted rating of item $i$ by user $u$ without considering their bias can be roughly viewed as the average similarity between item $i$ and item $j$, where $j \in I(u)$ (similarly to the factored item similarity model (FISM) proposed in \cite{FISM2013}).
	
	
	\noindent
	\textbf{Optimization.}\quad 
	The model parameters are learned by applying gradient descent (GD) across all users and items in the training set as follows:
	{\small
		\begin{align}
			&{ \frac{\partial O_{UR}}{\partial b_{u}}=\sum_{i \in R_{u}}e_{ui}+\lambda b_{u}} \label{eq:User-REG GD1}\\
			& { \frac{\partial O_{UR}}{\partial b_{i}}=\sum_{u \in U_{i}}e_{ui}+\lambda b_{i}} \label{eq:User-REG GD2}\\
			& { \frac{\partial O_{UR}}{\partial P_{u}}=\sum_{i \in R_{u}}e_{ui} Q_{i} +\beta(P_{u}-|I(u)|^{-1}\sum_{j\in I(u)}Q_{j})+\lambda P_{u}}  \label{eq:User-REG GD3}\\
			& \frac{\partial O_{UR}}{\partial Q_{i}}=\sum_{u \in U_{i}} e_{ui}P_{u}+\lambda Q_{i} \label{eq:User-REG GD4}
		\end{align}
	}%
	where  $e_{ui}=\tilde{r}_{ui}-{r}_{ui}$ and $U_{i}$ is the set of users that have rated item $i$. The pseudocode for learning UserReg  is presented in Algorithm \ref{alg:UserReg_alg}, where $\gamma$ is the learning rate.

	\begin{algorithm}[htp]
		\SetAlgoLined
		
		\caption{Parameter updating for UserReg.}
		\While{$ O_{UR} $ \textit{not converged}  }{
			compute the gradients according to \eqref{eq:User-REG GD1}-\eqref{eq:User-REG GD4}\\
			$b_{u} \leftarrow b_{u}-\gamma \frac{\partial O_{UR}}{\partial b_{u}}, u=1...m $\\
			$b_{i}\leftarrow b_{i}-\gamma \frac{\partial O_{UR}}{\partial b_{i}}, i=1...n $\\
			$P_{u}\quad\leftarrow  \quad P_{u}-\gamma \frac{\partial O_{UR}}{\partial P_{u}}, u=1...m$\\
			$ Q_{i}\quad\leftarrow \quad Q_{i}- \gamma \frac{\partial O_{UR}}{\partial Q_{i}}, i=1...n$\\
			
		}
		
		\label{alg:UserReg_alg}
	\end{algorithm}
	
	\noindent
	\textbf{Computational Complexity Analysis.}\quad 
	The main computational cost in each iteration of learning UserReg depends on the computational cost for computing the objective function and the gradients. Let $m$ be the total number of users and {\small $\overline{r}$} the average number of ratings for each user. Then, {\small $|R|\approx m\times \overline{r}$} and the computational complexity of the objective function for UserReg is {\small $\mathcal{O}(m \overline{r}d+m \overline{r}\overline{l}d)$}, where $\overline{l}$ is the average number of items a user is interested in. The complexity for computing the gradients according to ~\eqref{eq:User-REG GD1}-\eqref{eq:User-REG GD4} is {\small $\mathcal{O}(m \overline{r})$, $\mathcal{O}(m \overline{r})$, $\mathcal{O}(m \overline{r}d+m \overline{l}d)$}, and {\small $\mathcal{O}(m \overline{r}d)$}, respectively. The computation complexity of UserReg is {\small $\mathcal{O}(m \overline{r}d+m\overline{l}d)$} for one iteration. Due to {\small $\overline{r}, \overline{l},d  \ll m$}, the overall computational complexity is linear with respect to the number of users. 
	
	\noindent
	\textbf{Comparison with Other Similar Baselines.}\quad
	The models that are closely related to UserReg are SVD++ \cite{SVD++2008} and FISM \cite{FISM2013}.
	Compared to SVD++, UserReg: (1) predicts the rating of $ r_{ui} $ based on not only user $ u $, item $ i $ and the set of items that $ u $ has rated, but also considers the explicit ratings of items that has been rated by $ u $ and filters out the items that $ u $ dislikes; (2) incorporates the influence of items that a user liked as a regularization term, such that the prediction function remains the same with BiasedMF (equation (\ref{eq:BiasedMF})), which is more computationally efficient for rating prediction; and (3) it is more computational efficient in training, as the training computational complexity of SVD++ is linear with respect to the number of ratings \cite{SVD++2008} while in UserReg it is linear as regards the number of users. 
	Despite the fact that FISM and UserReg are designed for different recommendation tasks (item ranking vs. rating prediction), their key difference is the matrices they factorize. FISM is an item-oriented model that learns the item latent factors by factorizing the item similarity matrix, while UserReg learns latent factors for both users and items by factorizing the user-item rating matrix and  regularizing the user presentations by the set of items user likes.

	\section{PERFORMANCE COMPARISON }
	
	\subsection{Datasets}
	We evaluated the performance of UserReg using two extensively studied Movielens benchmark datasets (ML-100K and ML-1M)\footnote{http://grouplens.org/datasets/movielens/} and two datasets that are widely used for sparsity and cold start problems: FilmTrust \cite{guo2016TrustSVD} and the Yelp\footnote{http://yelp.com/dataset-change/}. FilmTrust contains both user ratings of items and social relations between users. The Yelp dataset, from the business domain, contains user ratings on businesses and attribute information of users and businesses. Main characteristics of the four datasets are shown in Table \ref{tab:dataset}. 

	\begin{table}[htp] 
		\caption{Descriptive statistics of the rating data in each dataset.}
		\label{tab:dataset}
		\centering
		\begin{tabular}{lrrrl}
			\toprule
			\bfseries Dataset & \bfseries \#Users & \bfseries\#Items & \bfseries \#Ratings & \bfseries Density\\
			\midrule
			ML-100K &	943&	1,682 &	100,000 & 6.31\%\\
			ML-1M&	6,040	&3,952 & 1,000,209 & 4.19\%\\
			FilmTrust&	1,508&	2,071	& 35,497 & 1.14\% \\
			Yelp  &16,239&	14,284	& 198,379 & 0.08\%\\
			\bottomrule
			
		\end{tabular}
	\end{table}
	
	Following \cite{AutoRec2015,NNMF2015,Zhang2020IGMC}, the performance of rating prediction is measured using the \textit{Root Mean Square Error} (RMSE): 
	$ {\footnotesize \sqrt{\frac{1}{|Z|}\sum_{(u,i) \in Z}(r_{ui}-\tilde{r}_{ui})^{2}}} $,
	where ${\textstyle Z }$ is the set of observed ratings in the test set. A smaller value of RMSE indicates better predictive accuracy of the corresponding  model.
	\begin{table}[t]
	\centering
		\caption{The RMSE resutls on Movielens-100K and ML-1M \textit{w.r.t.} different training test splits. (The first group are baselines, whereas the second group are newly proposed models, all are shown in ascending order by year of publication.\textit{ ct.} indicates whether content information is needed for the corresponding model.)}
		
		\subfloat[ML-100K]{
			\centering
			\label{tab:ml_100k}
			\begin{tabular}{llllll}
				\toprule
				Model  & 90/10 & ct. & Model  & 80/20 & ct.\\
				\midrule
				PMF  & 0.909 & no & PMF  & 0.918 & no  \\
				BiasedMF   & 0.908 & no& BiasedMF                & 0.917 & no    \\
				SVD++      & 0.903 & no  &SVD++                   & 0.914 & no   \\\hline
				NNMF (3HL) & 0.907 & no & GRALS     & 0.945 & yes   \\
				AutoSVD  & 0.901 & yes &mSDA-CF 	 &\textbf{0.904} & yes  \\
				AutoSVD++  & 0.904 & yes &  GC-MC  & 0.910 & no  \\
				NFM       & 0.910 & no & sRGCNN             & 0.929 & yes  \\
				GraphRec   & \textbf{0.898} & no  &IGMC   & 0.905 & no   \\
				\hline
				UserReg   & 0.901  & no & UserReg  & 0.906  & no   \\
				\bottomrule
			\end{tabular}
		}
	\hfill
		\subfloat[ML-1M]{
			\label{tab:ml_1m}\centering
			\begin{tabular}{llll}
				\toprule
				Model    & 90/10 & 50/50 & ct.\\
				\midrule
				PMF                                  & 0.871 & 0.879  & no  \\
				BiasedMF                        & 0.851 & 0.876 & no  \\
				SVD++                             & 0.850 & 0.874 & no  \\\hline
				NNMF (3HL)          & 0.846 & -& no  \\
				I-AutoRec  & 0.831 & -& no  \\
				CF-NADE (2HL)       & \textbf{0.829} &- & no  \\
				AutoSVD                       & 0.864 & 0.877 & yes \\
				AutoSVD++                     & 0.848 & 0.875 & yes \\
				NFM                          & 0.858 & 0.881 & no  \\
				GC-MC          & 0.832 &   -    & no \\
				GraphRec        & 0.845 & \textbf{0.862} & no  \\
				IGMC                           & 0.857 &   -    & no  \\\hline
				UserReg                        & 0.845 & 0.872  & no  \\

				\bottomrule	
		\end{tabular}}
			\vspace{-8pt}	

	\end{table}

	\subsection{Experiments}\label{sec:compare_study}
	Experiments were conducted in two groups: (1) on the Movielens datasets, as they are more general benchmarks used for various kinds of recommendation approaches; (2) on FilmTrust and Yelp, which are usually used for attribute-aware recommendations. For each dataset, we used experimental settings, including data splits, employed by the majority of previous models that have used that dataset. For the state-of-the-art models, we directly used the results reported in  the corresponding papers using the same experimental settings. For the baselines (PMF, BaisedMF, SVD++), as the results reported vary between different works with the same experimental settings, we obtained the results with fine-tuned parameters using the LibRec library \cite{guo2015librec}. For fair comparison, we set the number of latent factors in all MF-based models to $ 10 $. Each experiment was conducted five times and the average RMSE was taken as the final result.
	\subsubsection{On Movielens}
	 In addition to the three baselines, we also compared UserReg against multiple state-of-the-art rating prediction models.
	  GraphRec\cite{Graphrec2019} and IGMC\cite{Zhang2020IGMC} are the two most recently proposed CF models for rating prediction task using ML-100K and ML-1M. Based on \cite{Graphrec2019,Zhang2020IGMC}, we found a richer set of rating prediction models that present state-of-the-art results, including (1) graph neural networks (GNNs) based ones: GRALS \cite{NIPS2015_GRALS}, GC-MC \cite{berg2017GC_MF}, sRGCNN \cite{sRGCNN2017} and (2) neural network based CF models: mSDA-CF \cite{mSDA_CF2015}, NNMF (3HL) \cite{NNMF2015}, AutoSVD \cite{AutoSVD++2017}, AutoSVD++ \cite{AutoSVD++2017}, NFM \cite{he2017NFM}, CF-NADE \cite{CF-NADE2016}, I-AutoRec \cite{AutoRec2015}.
	 Following the experimental settings employed by the majority of these models, we used 90/10 and 80/20 train/test splits on ML-100K, and 90/10 and 50/50 train/test splits on ML-1M.

	For ML-100K, we used the hyper-parameters- $\lambda=0.1$ and  $\beta = 12$ for both data splits. For ML-1M, we used $\lambda=0.1$ and  $ \beta=8$ for the 90/10 train/test split, and $\lambda=0.1, \beta=5$ for the 50/50 split. The results are depicted in Table 2,
	where the best performance figures are shown in bold (results of recent models were taken from  \cite{AutoRec2015,AutoSVD++2017,Zhang2020IGMC,Graphrec2019}).


	According to the results, UserReg outperforms the three baselines on both datasets and achieves performance competitive with the state-of-the-art models which are either complicated hybrid CF models or use content information. On ML-100K, UserReg outperforms most of the recent models, except  mSDA-CF, GraphRec and IGMC, among which mSDA-CF use additional content information, and GraphRec and IGMC are based on GNNs. On ML-1M, UserReg cannot catch up with some of the NN-based models, such as I-AutoRec, CF-NADE, but still outperforms more than half of the remaining models. In particular, on both datasets UserReg achieves better performance than AutoSVD and AutoSVD++, which are powerful hybrid models \cite{Graphrec2019} that integrate MF with a contractive auto-encoder to tackle data sparsity in CF. 
	
	We also noticed that baseline results we obtained are better than those reported in most of the aforementioned recent works using the same experiment settings, which indicates that the baselines are not properly fine-tuned and the results in these works are suboptimal. Moreover, baseline results in more recent works are usually directly taken from previous studies, e.g. \cite{Graphrec2019,Zhang2020IGMC} that used results for PMF and BiasedMF taken from \cite{AutoSVD++2017}. However, the baselines in previous works (e.g. \cite{AutoSVD++2017}) were not properly fine-tuned, leading to the propagation of suboptimal baselines. This aligns with the findings stated in \cite{WorryingAnalysis2019} that the baseline results reported in previous works should be used with care.   

	\subsubsection{On FilmTrust and Yelp}\label{sec:film_yelp}
	As FilmTrust and Yelp are popular datasets for attribute-aware CF studies, besides baselines, we compared the performance of UserReg to some recently reported hybrid MF-based models that use additional content information. TrustSVD \cite{guo2016TrustSVD}, which is one of the most influencing trust-enriched MF. Additionally, we choose three more popular social relation enriched MF-based models, i.e. RSTE \cite{ma2009RSTE}, SoReg \cite{SoReg2011}, and TrustMF \cite{TrustMF2017}, for performance comparison on FilmTrust. 
	Yelp is widely used in studies on heterogeneous information networks (HINs). Besides HERec \cite{HERec2019}, which is the state-of-art HIN-based MF model, three other MF-based models that use HINs to construct context information were chosen for comparison on Yelp, namely HeteMF \cite{HeteMF2013}, SemRec \cite{SemRec2015} and DSR \cite{DSR2016}.
	We set the train/test split as 80/20 on both datasets. Results on FilmTrust are taken directly  from \cite{guo2016TrustSVD} and results on Yelp are taken from \cite{HERec2019}.
	For UserReg, we used $\lambda=0.1$ and $ \beta=1$ on FilmTrust, and $\lambda=0.1$ and $ \beta=10$ on Yelp. The obtained RMSE results are reported in Table \ref{tab:filmtrust}.
	
	\begin{table}[htp]
		\caption{The RMSE results on FilmTrust and Yelp for different models, shown in ascending order by year of publication.}
		\label{tab:filmtrust}
		\centering
		\begin{tabular}{llll}
			\toprule
			Model & FilmTrust & Model & Yelp\\
			\midrule
			PMF      & 0.968 & PMF      & 1.482 \\
			BiasedMF & 0.804 & BiasedMF & 1.096 \\
			SVD++    & 0.802 & SVD++    & 1.097 \\\hline
			RSTE     & 0.835 & HeteMF  & 1.280 \\
			SoReg    & 0.838 & SemRec   & 1.177 \\
			TrustSVD  & \textbf{0.789} & DSR      & 1.201 \\
			TrustMF   & 0.819 & HERec    & 1.112 \\\hline
			UserReg  & 0.798 & UserReg  & \textbf{1.041}\\
			\bottomrule
			
		\end{tabular}
	\end{table}
	
	The major findings are: (1) UserReg performs consistently better than the baselines on both datasets; (2) UserReg outperforms all state-of-the-art models (except TrustSVD on FilmTrust), despite the fact that it does not use any content information; (3) BiasedMF and SVD++ show better performance than most of the recently proposed models, and to the best of our knowledge, they have not been used as baselines for comparative study in the HIN-based research. 

		\begin{figure}[t]
	\centering
	\subfloat[][]{
	\begin{tikzpicture}
        \node (img) {\includegraphics[width=36mm,height=36mm]{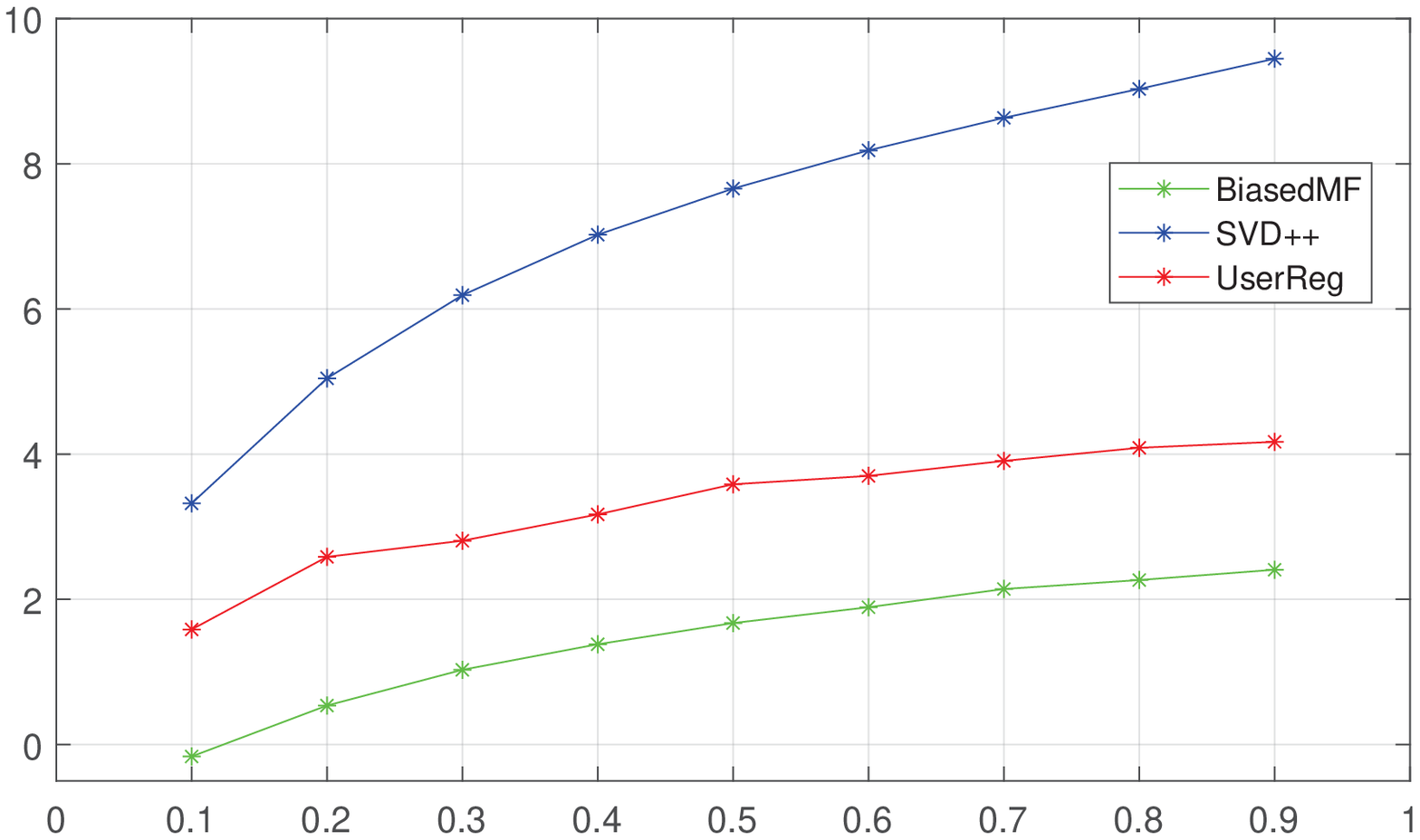}\label{fig:tiemcompare}};
        \node[node distance=0cm, yshift=-2.1cm] {\ninept training ratio};
        \node[node distance=0cm, rotate=90, anchor=center,yshift=2cm] {log (traning time)};
 \end{tikzpicture}}
	\subfloat[][]{
	\begin{tikzpicture}
	\node {\includegraphics[width=38mm,height=36mm]{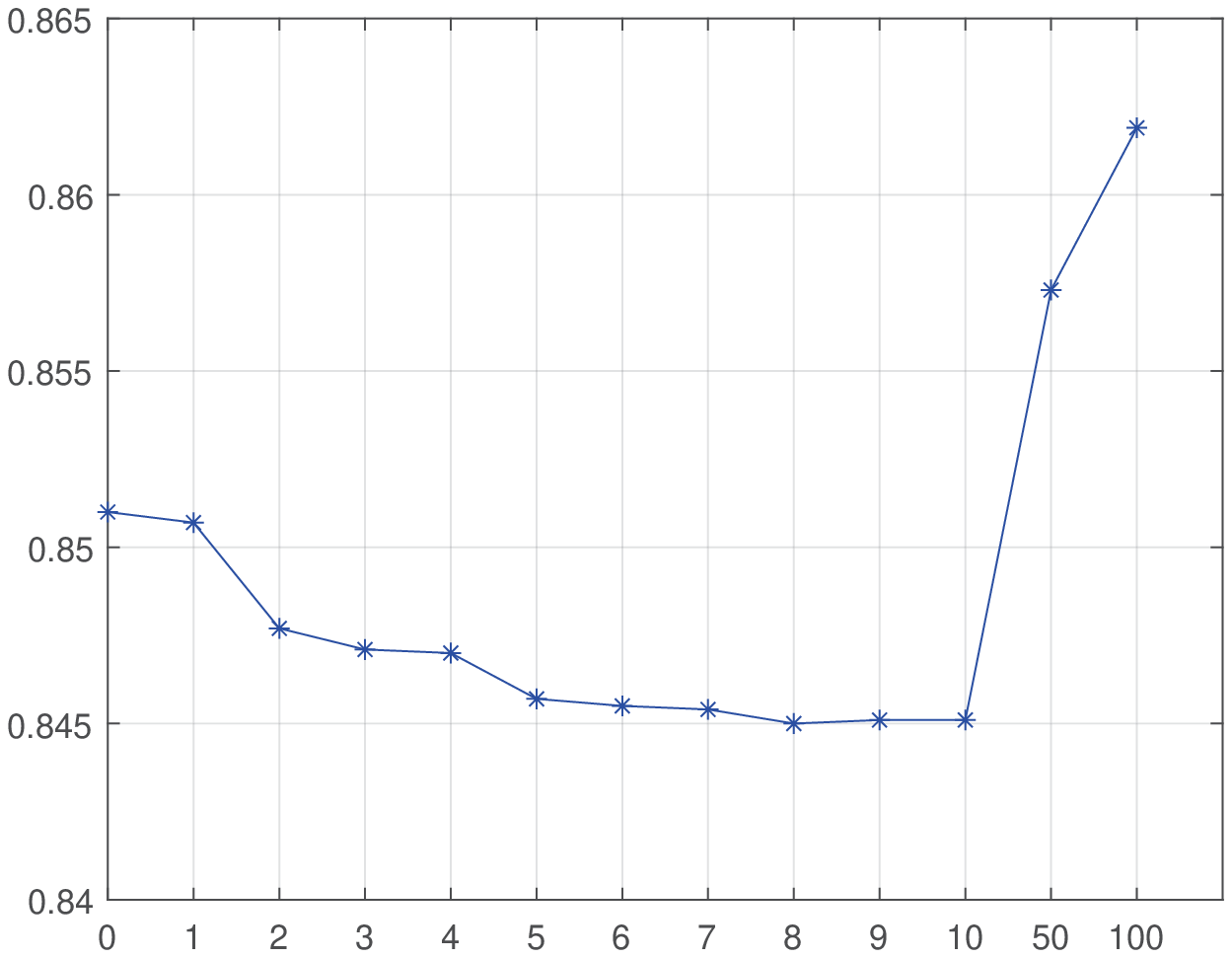}\label{fig:beta}};
	\node[node distance=0cm, yshift=-2.1cm] {\ninept $\beta$};
    \node[node distance=0cm, rotate=90, anchor=center,yshift=2cm] {RMSE};
\end{tikzpicture}
	}
	\caption{(a) A comparison of execution time on ML-1M with $d=5$. (b) The impact of parameter $ \beta $ on RMSE in UserReg.}
	\label{fig:experiment_last}
	\vspace{-8pt}

\end{figure}

	\subsection{Effectiveness vs. Efficiency}\label{sec:scalability}
	Figure \ref{fig:tiemcompare} shows the training time comparison (on a log scale) between BiasedMF,  SVD++ and UserReg, carried out on ML-1M.
	The execution time required by UserReg is significantly lower than that of SVD++, despite the latter being considered a strong baseline for rating prediction \cite{rendle2019difficulty}. Compared to BiasedMF and SVD++, UserReg provides a better trade-offs between recommendation effectiveness and efficiency.
	%

	\subsection{Impact of User Feedback Influence}
	The parameter $ \beta $ plays an important role in UserReg by controlling the impact of user feedback. In the extreme case when $ \beta = 0  $, UserReg equals BiasedMF and the user feedback has no influence on the predicted rating. When $ \beta $ is set to a relatively large value, the learning process is dominated by the influence of the items liked by the user. In order to analyse the impact of $ \beta $, we exprimented with its values in the range $(0, 100)$ on ML-1M with a 90/10 data split. Figure \ref{fig:beta} shows the impact of $\beta$ on RMSE. With the increase of $ \beta $, the prediction accuracy initially improves but later degrades, whereby the best performance is achieved for $\beta = 8$. Similar trends were found on other experimental settings and other datasets described. This indicates that the proper incorporation of user feedback  helps improve the recommendation accuracy.  
	
	\section{Conclusion}
	
	This paper has proposed a simple MF-based model, UserReg, that outperforms the fine-tuned baselines considered and achieves competitive results when compared to other more computationally complex state-of-the-art models. UserReg has the potential to be used as a strong baseline in the future development of rating prediction recommendations. In addition, findings have been presented that some of the baseline results reported in a number of newly published works are suboptimal. Moreover, some recent MF-based models have not been compared against strong baselines (BiasedMF and SVD++) as otherwise would be simply outperformed by them. The conclusion is that baselines used for performance  comparison should be chosen and fine-tuned with care in the future CF development.
	

	\vfill\pagebreak

	\bibliographystyle{IEEEbib}
	\bibliography{userReg}
	
\end{document}